\documentclass{aastex}
\include{epsf}
\usepackage{emulateapj5}

\shorttitle{The Chemo-Dynamical Evolution of Elliptical Galaxies}
\shortauthors{D. Kawata}

\begin{document}
\title{ 
Effects of SNe II and SNe Ia Feedback on the Chemo-Dynamical Evolution
of Elliptical Galaxies
}
\author{Daisuke Kawata}
\affil{Astronomical Institute,Tohoku University, Sendai,
980-8578, Japan}
\affil{ and }
\affil{ Centre for Astrophysics and Supercomputing,
 Swinburne University of Technology, Hawthorn, VIC 8122, Australia }
\email{dkawata@mania.physics.swin.edu.au}

\begin{abstract}

 We numerically investigate the dynamical and chemical processes
of the formation of elliptical galaxies in a cold dark matter (CDM) universe,
in order to understand the origin of the mass-dependence of
the photometric properties of elliptical galaxies.
Our three-dimensional TREE N-body/SPH
numerical simulations of elliptical galaxy formation
take into account both Type II (SNe II) and 
Type Ia (SNe Ia) supernovae (SNe)
and follow the time evolution of the abundances of several chemical
elements (C, O, Ne, Mg, Si, and Fe). 
Moreover we compare different strengths of SNe feedback.
In combination with stellar population synthesis,
we derive the photometric properties of simulation end-products,
including the magnitude, color, half-light radius, and
abundance ratios, and compare them with the observed scaling
relations directly and quantitatively.
We find that the extremely strong influence of SNe
is required to reproduce the observed color-magnitude relation (CMR),
where we assume each SN yields energy of $4\times10^{51}$ ergs
and that 90\% of this energy is ejected as kinetic feedback.
The feedback affects the evolution of lower mass systems more strongly
and induces the galactic wind by which a larger fraction of gas is 
blown out in a lower mass system. Finally higher mass systems 
become more metal rich and have redder colors than lower mass systems.
We emphasize based on our simulation results that the galactic wind
is triggered mainly by SNe Ia rather than SNe II.
In addition we examined the Kormendy relation,
which prescribes the size of elliptical galaxies, and
the [Mg/Fe]--magnitude relation, which provides a strong constraint
on the star formation history.  

\end{abstract}

\keywords{galaxies: elliptical and lenticular, cD
---galaxies: formation---galaxies: evolution
---galaxies: stellar content }

\section{Introduction}
\label{sintro}

Despite the large variety in the internal dynamics and 
structures, elliptical galaxies follow the various scaling
relations. These relationships are expected to contain valuable information
about physical processes of their formation.
For example, colors of elliptical galaxies strongly correlate
with their luminosities \citep[e.g.,][hereafter BLE92a and BLE92b]
{as72, apf81, ble92a, ble92b}.
This relation is usually called the color-magnitude relation (CMR).
The CMR is conventionally interpreted
as an effect of the galactic metallicity
increasing with the mass \citep[e.g.,][]{sf73, ad84},
although the age difference also contributes on the CMR possibly
\citep{wtf96}.
This interpretation is supported by line index analysis of galaxies
\citep{kd98, hk98, tkbcs99, hk00}
and slopes of CMRs observed in the intermediate redshift
($z\sim0.8$) clusters \citep{ka97}.
The mass-metallicity sequence
can be naturally explained by the galactic wind scenario \citep{rl74}
and reproduces the CMR very well,
according to the previous studies employing the evolutionary method
of population synthesis \citep[e.g.,][]{ay87, mt87, bg97, tcbmp98}.
In this scenario, galactic wind is induced progressively
later in more massive galaxies owing to deeper potential
and stellar populations in brighter galaxies
are much more enhanced in heavy elements, leading to redder colors.
However, these studies assumed a simple galaxy model
which ignored the internal structure and complex
star formation history in a forming galaxy.

Recently developed semi-analytic model \citep{wr87, wf91}
makes it possible to describe some complex star formation histories,
such as a star-burst induced by a coalescence of galaxies.
This model is based on the hierarchical clustering scenario
expected in the cold dark matter (CDM) cosmology.
The formation and merging histories of dark matter halos
are deduced by an extended Press-Schechter theory
\citep{rb91, bcek91} or a cosmological N-body simulation.
The model presumes that gas initially cools into
a disk and forms stars inside the halo.
A spheroidal stellar system is assumed to be formed
by a major merger of two disk galaxies \citep{kwg93, bcf96}.
\citet{kc98} showed by taking account of chemical enrichment
that the CMR of elliptical galaxies is reproduced in their
semi-analytic model with the assumption that
the supernova feedback is much efficient.
Although the recent semi-analytic models seem to succeed in
explaining the observational properties of galaxies at various redshifts,
they inevitably assumed a phenomenological model involving
a number of parameters to describe
the processes of radiative cooling, star formation
and feedback within the galactic halo.
Moreover, structures of the end-products could not be discussed,
because there is no information about the dynamical history
of gas and star components.

Numerical simulations are powerful tools
to treat the complex physical processes
in galaxy formation self-consistently.
 In recent works using three dimensional numerical simulations
which calculate the dynamical and chemical evolution self-consistently,
photometric properties of end-products were investigated
in combination with the stellar population synthesis and
compared to the observations directly
\citep[e.g.,][hereafter K99, K01]{bs98, bs99, csf98, myn99, sn99, ksw00a,
ksw00b, ns00, dk99, dk01}.
In order to understand the connection between
the observed scaling relations and the physical processes of the formation 
of elliptical galaxies, we investigate the dynamical and chemical
evolution of elliptical galaxies
in a cold dark matter (CDM) universe using the numerical simulation.
In this paper we focus on understanding the origin of the
CMR, i.e., how and what physical processes control the mass
dependence of the photometric properties. To this end,
we adopt the elliptical galaxy formation model of K99.
This model is based on a semi-cosmological model proposed by \citet{kg91}
and treats a collapse of a top-hat
over-dense sphere which initially follows a Hubble flow expansion and has
a solid-body rotation as an effect of the external tidal field.
In addition this sphere includes small-scale density fluctuations
expected in a CDM universe.
The galaxy is built by merger of small clumps
formed due to the small-scale fluctuations, i.e.,
a hierarchical clustering.
K99 studied the evolution of a seed galaxy which
has a slow rotation corresponding to a spin-parameter of $\lambda=0.02$
and showed that the end-product reproduces the observed
properties of elliptical galaxies very well.
Cosmologically, a small spin-parameter is preferred
in a galactic halo collapsing at a high redshift \citep{hp88}.
The reason for this is that higher peaks have a shorter collapse time;
thus, they have less time to get spun up by the environmental tidal force.
Moreover, a significant fraction of elliptical galaxies
are considered to be the systems which had collapsed at high redshifts,
according to the zero-point and tightness of the CMR
at the intermediate redshift \citep[e.g.,][]{bkt98, vffik00}.
It is thus a natural supposition that some elliptical galaxies
were formed in a halo collapsing at a high redshift and rotating slowly
\citep[e.g.,][]{bfpr84}.
Although K99 studied the dynamical and chemical evolution of
a galaxy with a fixed mass, we here examine
the photometric properties of the end-products with different masses.
We pay a particular attention to the supernovae feedback,
which is one of the most difficult processes to model in the
present-day numerical simulations, but considered to be
a crucial determinant of the nature of the stellar population
of the end-products.
Therefore, we examine how the difference in the strength of the feedback
affects the final CMR.
In addition our numerical code 
takes account of the chemical and dynamical feedback
of both Type II (SNe II) and Type Ia (SNe Ia) supernovae (SNe)
and follows the evolution of the abundances of several chemical
elements (C, O, Ne, Mg, Si, and Fe).
We evaluate relative importance of two types of SNe, SNe II and SNe Ia,
on the dynamical and chemical evolution.
Since our numerical simulations naturally
provide the information about the structure of end-products
with enough spatial resolution,
we compare simulation results
not only with the observed CMR but also with the Kormendy relation, 
which prescribes the size of elliptical galaxies as a function
of their luminosity. 
Due to self-consistent calculation of evolution of chemical elements abundances,
we can also study the mass dependence of abundance ratios.
\citet{hk98, hk00} studied early-type galaxies in the Fornax
cluster, using the line indices analysis for high-quality data.
He found that Fornax ellipticals are mainly old and also discovered
a strong relation between [Mg/Fe] and the velocity dispersion
\citep[see also][]{tfwg00b}. \citet{ij99}
 studied Coma elliptical and obtained
a strong correlation between [Mg/Fe]
and not only the velocity dispersion but also the luminosity.
Hereafter we call the correlation between [Mg/Fe] and the luminosity
``[Mg/Fe]--magnitude relation''.
Because Mg and Fe are mostly produced by SNe II and
SNe Ia respectively and because SNe Ia have a longer delay
than SNe II after formation of stars, these correlations give strong
constraints on the star formation history of elliptical galaxies
\citep{mpg98, tgb99}.
We examine the [Mg/Fe]--magnitude relation for the simulation
end-products and compare it with the observation.

 The plan for the remainder of this paper is as follows.
The numerical method and the model used in this paper
are described in Sections \ref{scode} and \ref{smodel}.
We show the procedure to derive the photometric properties
of the simulation end-products in Section \ref{sdataa}.
In Section \ref{sgsrs}, results of numerical simulations are presented
and are compared with the observed global scaling relations.
Our discussion and conclusions are given in Section \ref{sdisc}.

\section{The Code}
\label{scode}

 Numerical simulations are performed using 
an update version of the code in K99
so that SNe Ia feedback is taken into account.
Because we have already described the details of the code in K99,
here we describe mainly the modelings of star formation and their
feedback which are revised.
Our code is essentially based on the TreeSPH \citep{hk89, kwh96},
which combines the tree algorithm \citep{bh86}
for the computation of the gravitational forces with the smoothed
particle hydrodynamics \citep[SPH:][]{l77,gm77}
approach to numerical hydrodynamics.
The dynamics of the dark matter
and stars is calculated by the N-body scheme, and the
gas component is modeled using the SPH.
It is fully Lagrangian, three-dimensional, and highly adaptive in space
and time owing to individual smoothing lengths and
individual time steps. Moreover, it self-consistently
includes almost all the important physical processes
in galaxy formation, such as self-gravity, hydrodynamics,
radiative cooling, star formation, supernova feedback, and
metal enrichment.

The radiative cooling which depends on the metallicity \citep{tbh92}
is taken into account. The cooling rate of a gas with the solar metallicity
is larger than that for a gas of the primordial composition
by more than an order of magnitude.
Thus, the cooling by metals should not be ignored
in numerical simulations of galaxy formation
\citep{kh98,kpj00}.

\subsection{Star Formation}
\label{ssf}

 We model star formation using a method similar to
that of \citet{nk92} and \citet{kwh96}.
We use the following three criteria for star formation:
1) the gas density is greater than a critical density,
$\rho_{\rm crit} = 2 \times 10^{-25}\ {\rm g\ cm^{-3}}$,
i.e., $n_{\rm H} \sim 0.1 {\rm cm^{-3}}$, following \citet{kwh96}; 
2) the gas velocity field is convergent,
${\bf \nabla} \cdot \mbox{\boldmath $v$}_i < 0$; and 3) the 
Jeans unstable condition, $h/c_s>t_{\rm g}$, is satisfied,
here $h$, $c_s$, and $t_{\rm g} = \sqrt{3 \pi/16 G \rho_{\rm g}}$ 
are the SPH smoothing length,  
the sound speed, and the dynamical time of the gas respectively
(see K99 for details).
The Jeans condition was ignored in K99 and K01, 
because the gas whose density is higher than
$\rho_{\rm crit}$ usually satisfies this condition. 
Since the Jeans unstable condition might be important in the low mass
systems, we take it into account in this paper.

 When a gas particle is eligible to form stars,
its star-formation rate (SFR) is
\begin{equation}
 \frac{d \rho_*}{dt} = -\frac{d \rho_{\rm g}}{dt}
 = \frac{c_* \rho_{\rm g}}{t_{\rm g}},
\label{sfreq}
\end{equation}
where $c_*$ is a dimensionless SFR parameter
and $t_{\rm g}$ is the dynamical time, 
which is longer than the cooling timescale
in the region eligible to form stars.
This formula corresponds to the Schmidt law that
SFR is proportional to $\rho_{\rm g}^{1.5}$.
The value of $c_*$ controls the SFR
with respect to the local gas density, namely star formation efficiency.
When a large $c_*$ is assumed,
many stars can be formed well before the system collapses completely,
so that the effective radius of the final stellar system becomes large.
In this paper we assume for simplicity that $c_*$ is constant irrespective of 
the mass of the system and calibrated $c_*$ so that model A2 defined
in Section \ref{smodel} could reproduce
the observed Kormendy relation. Figure \ref{fmre} shows the results of
model A2 in the case of $c_*=1$ (cross) and $c_*=0.5$ (triangle just under the
cross) respectively. 
We can see that $c_*=1$ leads to a little larger
effective radius than the observed elliptical galaxies under the assumed
distance modulus (see Section \ref{sgsrs} for details). 
Although the difference in $c_*$ a little affects
the other photometric properties which we examine in this paper
(Fig.\ \ref{fcmr} and \ref{fmgfe}), finally we found that
this effect is quite small, compared to difference in the
strength of SNe feedback which we focus on in this paper.
Consequently we set $c_*=0.5$.
In addition the above critical gas density, $\rho_{\rm crit}$,
is higher than that in K99 and K01. We also carried out
a simulation with the same $\rho_{\rm crit}$ as K99
and found that this difference does not change the effective radius
in model A2, because the gas density becomes higher than these densities
very quickly once the system collapses. The difference
in $\rho_{\rm crit}$ within this range does not affect our results.

 We assume that the stars which are represented by a star particle
are distributed according to the \citet{s55} initial mass function (IMF).
The IMF by number, $\Phi(m)$, in each mass interval $dm$
is defined as
\begin{equation}
 \Phi(m) dm = Am^{-(1+x)} dm,
\label{imfeq}  
\end{equation}  
where $x=1.35$ is the Salpeter index (this corresponds
to an IMF by mass $\propto m^{-x}$) and the coefficient $A$
is determined by the normalization in the mass range
$M_{\rm l} \leq m \leq M_{\rm u}$.
We set $M_{\rm u}=60\ {\rm M_\odot}$ and $M_{\rm l}=0.4\ {\rm M_\odot}$
respectively.

\subsection{The Feedback}

 We take account of the energy feedback and metal enrichment 
to the surrounding gas by SNe. 
We consider here both SNe II and SNe Ia. 
The simulation follows the evolution of the abundances
of several chemical elements (C, O, Ne, Mg, Si, and Fe).

\subsubsection{Type II Supernovae}

 For simplification, we assume that each massive star ($>10\ {\rm M_\odot}$)
explodes as a type II supernova within the simulation time step
in which it was born (instantaneous recycling). 
From equation (\ref{imfeq})
the number of SNe II in a new-born star particle of $m_{\rm s}\ {\rm M_\odot}$ 
becomes $N_{\rm SNe II} = 9.26\times10^{-3} m_{\rm s}$.
The total stellar mass contributing to SNe II is
$M_{\rm SNe II} = 1.83\times10^{-1} m_{\rm s}\ {\rm M_\odot}$ and
the mass of $(M_{\rm SNe II}-1.4 N_{\rm SNe II})\ {\rm M_\odot}$
is returned to the the surrounding gas, where
the remnant mass is assumed to be $1.4\ {\rm M_\odot}$.
In this process, a portion of the metals
produced in the stars is also returned to the gas,
leading to chemical enrichment. We use the stellar yields derived
by \citet{nht97} who provide for SNe II the $i$-th heavy element
mass, $M_{i, {\rm SNe II}}(m)$, produced in a star of
 main-sequence mass of $m$.
Then the synthesized mass of SNe II in a new-born star particle
of $m_{\rm s}\ {\rm M_\odot}$ for each element is calculated by
\begin{equation}
 M_{i, {\rm SNe II}} = m_{\rm s}
  \int_{10\ {\rm M_\odot}}^{M_{\rm u}} M_{i, \rm SNe II}(m)
  m^{-(1+x)} dm
  \left/
 \int_{M_{\rm l}}^{M_{\rm u}} m^{-x} dm
  \right.
 \ {\rm M_\odot}.
\end{equation}
Table \ref{tblmet-1} shows $M_{i, {\rm SNe II}}$ for each element.
Here we assume that the heavy-element production from a star
of 10 ${\rm M_\odot}$ is negligible, following \citet{tny95}.

 These mass, energy, and heavy elements
are smoothed over the neighboring gas particles using the SPH
smoothing algorithm.
For example, when the $i$-th particle changes from gas to a star,
the increment of the mass of the $j$-th neighbor particle
due to explosion of the new-born star is given by
\begin{equation}
 \Delta { M_{\rm SN,{\it j}}}
  =  \frac{m_j}{\rho_{{\rm g},i}} { M_{\rm SN,{\it i}}}
  W(r_{ij},h_{ij}),
\label{msneq}
\end{equation}
where
\begin{equation}
 \rho_{{\rm g},i} = \langle \rho_{\rm g}(\mbox{\boldmath $x$}_i) \rangle
 = \sum_{j \neq i} m_j W(r_{ij},h_{ij})
\end{equation}
and $W(r_{ij},h_{ij})$ is an SPH kernel (see K99).

\subsubsection{Type Ia Supernovae}
\label{ssneia}

 Comparing the predicted nucleosynthesis products of SNe Ia and SNe II
with the solar abundances of heavy elements and their isotopes,
\citet{tny95} derived that the ratio of the total numbers of
SNe Ia to SNe II is $N_{SNe Ia}/N_{SNe II}=0.15$.
If the progenitor of SNe Ia is
a binary system one of which has a mass range of $m=3$-$8\ {\rm M_\odot}$,
$N_{\rm SNe Ia}/N_{\rm SNe II}$ is calculated by
\begin{equation}
 \frac{N_{\rm SNe Ia}}{N_{\rm SNe II}} =
 \frac{A_{\rm Ia} \int_{3\ {\rm M_\odot}}^{8\ {\rm M_\odot}} m^{-(1+x)} dm}
   {\int_{10\ {\rm M_\odot}}^{60\ {\rm M_\odot}} m^{-(1+x)} dm}.
\end{equation} 
Thus, we adopt $A_{\rm Ia} \sim 0.04$. Then
the number of SNe Ia in a new-born star particle of $m_{\rm s}\ {\rm M_\odot}$ 
becomes $N_{\rm SNe Ia} = 1.52\times10^{-3} m_{\rm s}$ and
the ejected mass is assumed to be $A_{\rm Ia}$ times the total mass of stars
whose mass is between 3 ${\rm M_\odot}$ to 8 ${\rm M_\odot}$, i.e., 
$M_{\rm SNe Ia} = 6.94\times10^{-3} m_{\rm s}\ {\rm M_\odot}$.
The nucleosynthesis prescriptions for SNe Ia are taken
from \citet{nin97} as shown in Table \ref{tblmet-1}.
If SNe Ia occur in a binary systems with the above mass range,
SNe Ia should have various lifetime
depending on the main-sequence lifetime of a secondary star.
Therefore, SNe Ia are expected to continuously occur for a long time
once they take place in the most massive binary system \citep[e.g.,][]{gr83}.
Moreover, most recently a new SNe progenitor scenario
in which SNe Ia rate depends on the metallicity was suggested
and it was shown that this scenario is suitable for
the explanation of the observed chemical evolution in the solar neighborhood
\citep{ktnhk98}.
However, we fix the lifetime of SNe Ia to 1.5 Gyr for simplicity,
following a result of \citet{ytn96}, who presented
that the mean lifetime of the progenitors of SNe Ia is about 1.5 Gyr,
based on the chemical evolution analysis in the solar neighborhood.

\subsubsection{The Energy Feedback}

 One of the most difficult and most critical processes to model
in galaxy formation simulations is the way in which the feedback from
SNe affects the surrounding gas. Unfortunately, there is no
well understanding of how it should be modeled.
In a pioneering SPH simulation by \citet{nk92}
the energy feedback was implemented as deposition of purely thermal
energy into the surrounding gas.
\citet{nk92} founds that this form of energy feedback
is rather inefficient; the high densities of typical star-forming
regions imply short cooling timescales that minimize the hydrodynamical 
effects of the feedback energy input. 
In real star-forming systems SNe are largely responsible for
the multi-phase structure of the interstellar medium \citep[e.g.,][]{mo77}.
Although they are associated with the cool
and dense gas clouds, they can deposit much of their energy in a
hot and low density wind. A seminal attempt at representing
this multi-phase interstellar medium (ISM) has been carried out
by \citet{ykkk97} and recently \citet{hp99}
have adapted the Yepes et al.\ model to SPH simulations.
However, they had to make a number of assumptions about the physics
of the ISM and their models are not truly multi-phase, because
different phases are treated as a dynamically single entity.
Another implementation assumed that the feedback region
around newly born stars evolves adiabatically until the multiple
SNe II phase ends \citep[e.g.,][]{mytn97,tc00}.
Although this model is somewhat artificial,
it can provide strong effects on the formation
of dwarf galaxies \citep{mytn97}.

While a number of models for energy feedback are proposed by previous
authors as mentioned above, we adopt here
the model proposed by \citet{nw93}.
This model assumes for simplicity that the energy produced by SNe affects only
the temperature and velocity field of the surrounding gas
and its effect is implemented by increasing 
the internal and kinetic energy of the gas neighbors of each star particle
by amount corresponding to the energy released by SNe.
Thus, when $i$-th new born star particle ejects
the energy of $E_{{\rm SN},i}$ and
deposits the energy of $\Delta E_{{\rm SN},j}$
in the $j$-th neighbor gas particle, the velocity of the $j$-th gas
is altered by $\Delta v_{{\rm SN},j}$, which is calculated from
\begin{equation}
f_v \Delta E_{{\rm SN},j} =
 f_v E_{{\rm SN},i} \frac{m_j}{\rho_{{\rm g},i}} W(r_{ij},h_{ij})
 = m_j (\mbox{\boldmath $v$}_j \cdot \Delta \mbox{\boldmath $v$}_{{\rm SN},j}
 +\frac{1}{2}\Delta v_{{\rm SN},j}^{2}),
\label{eqfvesn}
\end{equation}
where $\mbox{\boldmath $v$}_j$ is the relative velocity to the $i$-th star,
$\Delta \mbox{\boldmath $v$}_{{\rm SN},j}$ is a velocity perturbation
directed radially from the $i$-th star, and $f_v$ is an
input parameter which controls the fraction of the available energy
to perturb the gas velocity field.
The rest of the energy of the SNe, $(1-f_v) \Delta E_{{\rm SN},j}$,
contributes to increase the internal energy of $j$-th gas.
When the term of
$\mbox{\boldmath $v$}_j \cdot \Delta \mbox{\boldmath $v$}_{{\rm SN},j}$
is assumed to be zero approximately \citep{nw93},
the magnitude of the $\Delta \mbox{\boldmath $v$}_{{\rm SN},j}$
is then simply 
\begin{equation}
 \Delta v_{\rm SN,{\it j}}^{2}
  =  2\frac{f_v \Delta E_{\rm SN,{\it j}}}{m_j}.
\label{eqdvsn}  
\end{equation}
This model is consistent with the model of \citet{nk92}, in which $f_v=0$.
The amount of effects of the energy feedback on forming galaxies
depends quite strongly on the value of $f_v$.
It is known that the kinetic feedback affects the history of 
star formation more strongly 
than the thermal feedback which quickly dissipates
due to the radiative cooling in the high density region where stars can
form \citep{nw93}.
Thus, $f_v$ controls the strength of the effect of SNe.

 We assume that each SN yields energy of 
$\epsilon_{SN} \times 10^{51}$ ergs.
According to high resolution 1D simulations of SN remnant in \citet{tgjs98},
90 \% of an initial SN energy is lost in radiation in its early 
expansion phase, which is not resolved in our simulations.
However, an initial SN energy has not been established quantitatively yet.
Therefore, we consider the available SN energy as a free parameter.

 In our code, there are two parameters to control the strength
of the effect of SNe.
To examine the effect of the difference in the strength of 
the energy feedback, we carry out the simulations adopting
two extreme sets of these parameters, namely 
$(\epsilon_{SN},f_v)=(0.1,0)$ and $(\epsilon_{SN},f_v)=(4,0.9)$.
The former means a minimum feedback model, corresponding to
the result of \citet{tgjs98} when an initial SN energy $10^{51}$ ergs
is assumed.
The latter provides a feedback strong enough to reproduce the observed CMR 
as shown later.  

\section{The Model}
\label{smodel}

Using the above code, we calculate the following semi-cosmological model
which is almost the same as the one in K99.
We consider an isolated sphere, as a seed galaxy,
on which small-scale density fluctuations
corresponding to a CDM power spectrum are superimposed.
Here, we use Bertschinger's software COSMICS \citep{eb95}
in generating initial density fluctuations.
To incorporate the effects of fluctuations with longer wavelengths,
the density of the sphere has been enhanced and a rigid rotation
corresponding to a spin parameter, $\lambda$, has been added.
The initial condition of this model is determined
by the following four parameters: $\lambda$, $M_{\rm tot}$,
$\sigma_{\rm 8,in}$, and $z_{\rm c}$.
The spin parameter is defined by
\begin{equation}
 \lambda \equiv \frac{J|E|^{1/2}}{G M_{\rm tot}^{5/2}},
\end{equation}
where $J$ is the total angular momentum of the system,
$E$ is the total energy,
and $M_{\rm tot}$ is the total mass of this sphere, which is
composed of dark matter and gas;
$\sigma_{\rm 8,in}$ is the rms mass fluctuation in a sphere of radius
$8\ h^{-1}$ Mpc, which normalizes the amplitude of the CDM power spectrum;
$z_{\rm c}$ is the expected collapse redshift.
 If the top-hat density perturbation has an amplitude of
$\delta_i$ at the initial redshift, $z_i$, we obtain
$z_{\rm c} = 0.36 \delta_i (1+z_i)-1$ approximately
\citep[e.g.,][]{tp93}.
Thus, when $z_{\rm c}$ is given, $\delta_i$ at $z_i$ is determined.

K99 found that
the seed galaxy which has a slow rotation corresponding to $\lambda=0.02$
and the small-scale density fluctuations evolves into an elliptical-like
system using the similar numerical simulation to that of this paper.
This spin parameter is close to the minimum value possible
in the CDM universe, according to the results of N-body simulations
\citep{be87,wqsz92}.
In this paper we adopt this elliptical galaxy formation scenario.
The main purpose of this paper is to study
the evolution of seed galaxies with different masses.
In addition we consider that the strength of feedback is crucial
to give rise to mass-dependent evolution process.
Thus, we simulate the evolution of the seed galaxies
with the total mass between $8\times10^{12}$ and $1\times10^{11}$
${\rm M_\odot}$, using the two codes with different strength of 
the SNe feedback as mentioned above.
Parameters in each model are summarized in Table \ref{tblmps}.
Each model is labeled, for example, model 'A1', according to
the total mass and the strength of the SNe feedback.
Models A have $(\epsilon_{SN},f_v)=(0.1,0)$, whereas 
$(\epsilon_{SN},f_v)=(4,0.9)$ in models B.
The number in the model name represents the total mass of the seed galaxy.
Hereafter we refer to the mass sequence from model A1 (B1) to model A4 (B4)
simply as ``model A (B)''.
In all the models, we fix $\sigma_{\rm 8,in}=0.5$ and $z_c=3.5$.

 Our simulations assume a flat universe ($\Omega = 1$)
with a baryon fraction of $\Omega_{\rm b} = 0.1$
and a Hubble constant of $H_0 = 50\ {\rm km\ s^{-1}\ Mpc^{-1}}$.
We carry out each simulation using 9171 particles for gas and dark matter
respectively. The mass and spatial resolutions for each model
are shown in Table \ref{tblmps}.
We simulate the evolution of each model from $z_i=40$ to $z=0$.
The morphological evolutions of all the models are similar to the evolutions
seen in Figure 1 of K99 and K01. 
Finally, nearly spherical stellar systems 
are formed at $z=0$ in all the models.

\section{Data Analysis}
\label{sdataa}

 To compare simulation results with the observational data
of nearby elliptical galaxies directly,
we have to deduce photometric properties of simulation end-products
at $z=0$ for various models. We use the same procedures as K01.
Here we briefly describe the procedure to derive the photometric properties
from simulation results.
For this purpose, we show the analysis for model A2 as an example.

 In our simulations, the stellar particles contain the information about their
age and metallicity due to the self-consistent calculation
of the chemical and dynamical evolution.
By means of the population synthesis, we can derive the
photometric properties of the stellar system from this information.
Here, the spectral energy distribution (SED) of each stellar particle
is assumed to be that of a single stellar population (SSP)
that means a coeval and chemically homogeneous assembly of stars.
Since the observational data with which our results should be compared
provide the luminosity distribution projected to a plane,
we have to derive the projected distribution of SED
from the three dimensional distribution of stellar particles.
Finally, we obtain the projected images as shown in Figure 5 of K01.
Then the flux of each stellar particle is
smoothed using a gaussian filter with the filter scale of 1/4 of the
softening length of the stellar particle.
These images provide quite similar information to the imaging data obtained in
actual observations. Thus we can obtain various photometric properties from
these images in the same way as in the analysis of observational imaging data.
In the following analysis, we use the images similar to
the one displayed in Figure 5 of K01,
but employing a 1001 $\times$ 1001 pixel mesh
to span the squared region with 100 kpc on a side.

\vbox{
\begin{center}
\leavevmode
\hbox{
\epsfxsize=8.5cm
\epsffile{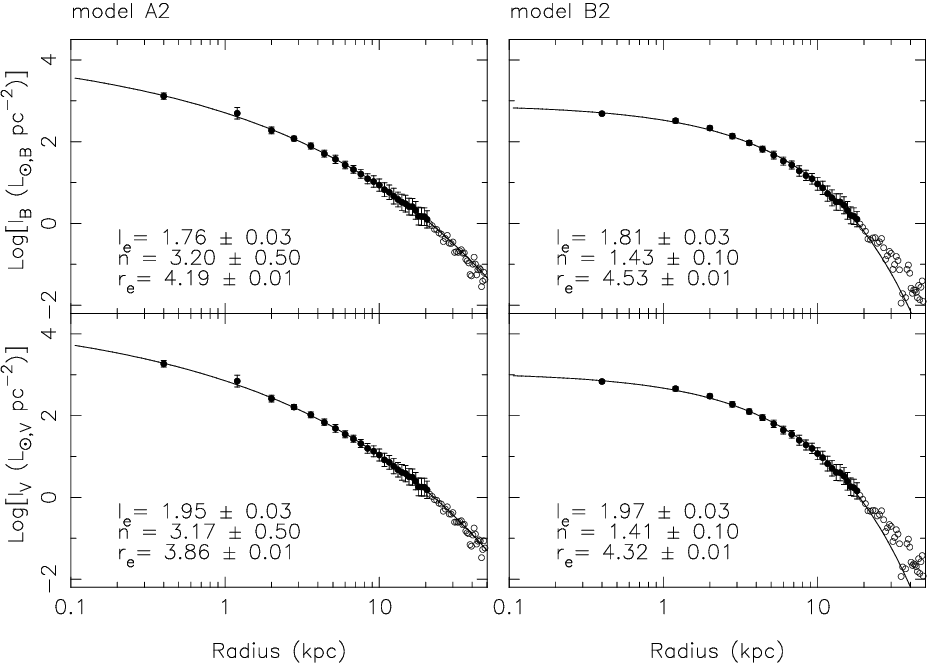}}
\figcaption[f1.eps]
{ The $B$ (upper panel) and $V$ (lower panel) band surface brightness profiles 
in the $x-y$ projections for models A2 (left) and B2 (right).
 The solid lines denote the $r^{1/n}$ law best-fit for
 the data plotted as the solid symbols. The fitting parameters
 are shown in the lower left corner of each panel.
 The error bars are shown only for the solid symbols and correspond to
 the standard deviation (see text for details).
\label{fa2sfb} }
\end{center}}

 In this paper, we use the data of SSPs of Kodama \& Arimoto 97 model
\citep{tk97,ka97}.
Kodama \& Arimoto 97 model supplies the database of SSPs with two types of IMF:
($x$, $M_{\rm u}$, $M_{\rm l}$) = (1.35, 60, 0.1) and
($x$, $M_{\rm u}$, $M_{\rm l}$) = (1.1, 60, 0.1)
in the definition of Section \ref{ssf}.
We adopt the data of SSPs with
the IMF of ($x$, $M_{\rm u}$, $M_{\rm l}$) = (1.35, 60, 0.1),
while we use the IMF of ($x$, $M_{\rm u}$, $M_{\rm l}$)
= (1.35, 60, 0.4) in the numerical simulations.
We confirmed that the photometric properties presented in
this paper are hardly changed, when we employ
another dataset of SSPs with the IMF of
($x$, $M_{\rm u}$, $M_{\rm l}$) = (1.1, 60, 0.1).
A shallower IMF has a similar fraction of massive stars
to the Salpeter IMF ($x=1.35$) whose lower mass limit, $M_l$, is
a little higher. Therefore, our use of the IMF of
($x$, $M_{\rm u}$, $M_{\rm l}$) = (1.35, 60, 0.1)
is justified.

 The left panels of Figure \ref{fa2sfb} shows the surface-brightness
profiles for model A2 
in the $B$ and $V$ bands obtained by setting the annuli of various radii
in each band image. The error bars show the standard deviation 
in each annulus. We set the width of each annulus 
to half the softening length.
Since the error is calculated in the flux, like
$\sigma_{f_{B}}^2=<f_{B}^2>-<f_{B}>^2$,
the error in the magnitude is defined, using the Taylor expansion, as
\begin{equation}
\sigma_{M_B}=2.5\left(\frac{\sigma_{f_{B}}}{<f_{B}>\ln (10)}
-\frac{\sigma_{f_{B}}^2}{2<f_{B}>^2 \ln(10)} \right).
\end{equation}
These surface brightness profiles are in excellent agreement
with the Sersic ($r^{1/n}$) law \citep{js68},
\begin{equation}
 I(r)=I_e 10^{\{-b_n[(r/r_{{\rm e}})^{1/n}-1]\}},
\label{eqserlaw}
\end{equation}
where we adopt $b_n=0.868n-0.142$, so that the effective radius, $r_e$,
equals the half light radius in the range $0.5\leq n \leq 16.5$
\citep{ccd93} and $I_e$ is the surface brightness at $r_e$.
This corresponds to the de Vaucouleurs ($r^{1/4}$) law
\citep{gd48}, when $n=4$.
The solid lines in Figure \ref{fa2sfb} show the best-fits
in applying equation (\ref{eqserlaw}).
In fitting, we use the data with $\mu_B < 27\ {\rm mag\
arcsec^{-2}}$ irrespective of the observed band.
The best-fit parameters are shown in the lower left corner
of each panel in Figure \ref{fa2sfb}.

\vbox{
\begin{center}
\leavevmode
\hbox{
\epsfxsize=8.5cm
\epsffile{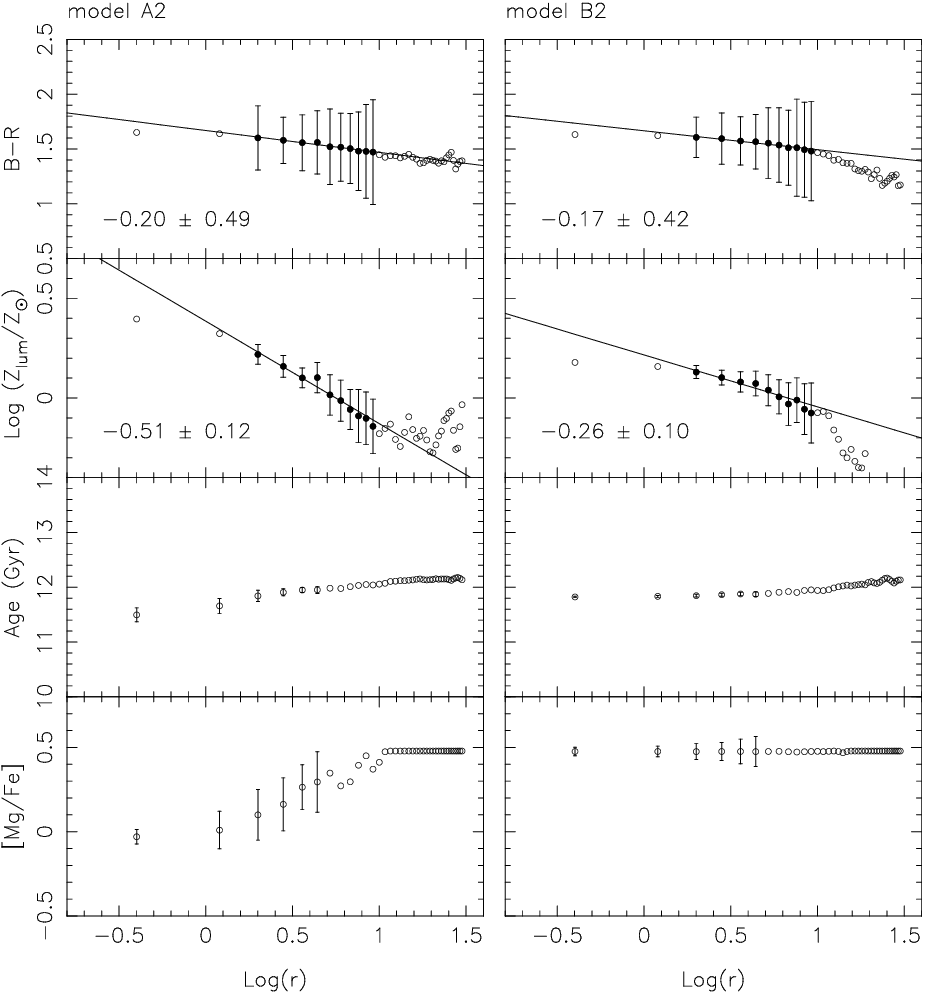}}
\figcaption[f2.eps]
{Color (top), luminosity weighted metallicity (2nd),
 luminosity weighted age (3rd), and [Mg/Fe] (bottom) gradients
 in the $x-y$ projection for models A2 (left) and B2 (right).
 The solid lines show the best-fit linear relations for
 the data plotted as the solid symbols.
 The gradients in best-fit lines
 are shown in the lower left corner of each panel.
 The error bars are shown only for the solid symbols 
 in top and 2nd panels and for the data of $r<5$ kpc in 3rd 
 and bottom panels.
\label{fa2grad}}
\end{center}}

 The left panels of Figure \ref{fa2grad} shows the color, metallicity, 
and age gradients for model A2.
The color gradients are obtained by setting annuli of
various radii in the $B$ and $R$ band images
and subtracting the $R$ band magnitude from the $B$ band magnitude
in each annulus.
We made the projected images also for the metallicity and age,
and obtained their radial profiles.
Points indicate the average value in each annulus
in the $x$-$y$ projections.
The width of each annulus is set to half the softening length.
The error bars show the standard deviation in each annulus.
The error in the color is written as
$\sigma_{B-R}=\sqrt{\sigma_{M_B}^2+\sigma_{M_R}^2}$.
The profiles of color and metallicity [defined as
$\log (Z/Z_\odot)$] are fitted by linear regression.
In fitting, we excluded the data at radii less than the softening
length and greater than the radius at which the $B$ band surface brightness
is $\mu_B=24.5\ {\rm mag\ arcsec^{-2}}$ (open symbols in
Fig.\ \ref{fa2grad}), because the inner region is affected by
the smoothing of gravitational forces whereas the number of particles
within an annulus is too small in the outer region.
The best-fit gradients are shown in the lower left corner
of each panel in Figure \ref{fa2grad} and also in Table \ref{tabglp}.
The data obtained from the simulation results
have large errors leading to large errors in the calculated gradients.
However negative gradients are clearly seen in colors
and metallicity, i.e., the color (metallicity) at the center is
redder (higher) than that in the outer region.
Although we show only a luminosity weighted metallicity
to compare with the observational data, 
the metallicity gradient weighted by the luminosity is
slightly steeper than that weighted by the mass in all the models.
It is known that a typical elliptical galaxy has
$\Delta(B-R)/\Delta\log(r)=-0.09\pm0.02$ \citep{pdidc90}
and $\Delta\log(Z/Z_\odot)/\Delta\log(r)=-0.30\pm0.12$
\citep[][and references therein]{ka00}.
Although the gradients obtained from the simulation results
have large errors, the median gradients
of both color and metallicity for model A2 are steeper
than those observed.
We can see that the color gradients are caused by the metallicity
gradients, because the age is equally old irrespective of the radius.
The radial profile of the age has a slightly positive gradient.
This positive gradient means that star formation persists
over longer time in the inner region than in the outer region,
due to dissipative infall of gas.
Since the residual gas is polluted
by past star formation, metal rich stars are formed in
the central region where the duration of star formation is long
(K99). This is the origin of the color and metallicity
gradients. Recent observations using the metal absorption lines
provide no (for ellipticals) or positive (for S0s) gradients
of the age \citep{hk98}. Since the age gradient
in Figure \ref{fa2grad} is too shallow, this gradient seems not to be
detected in the current observations using the absorption lines.

\vbox{
\begin{center}
\leavevmode
\hbox{
\epsfxsize=8.5cm
\epsffile{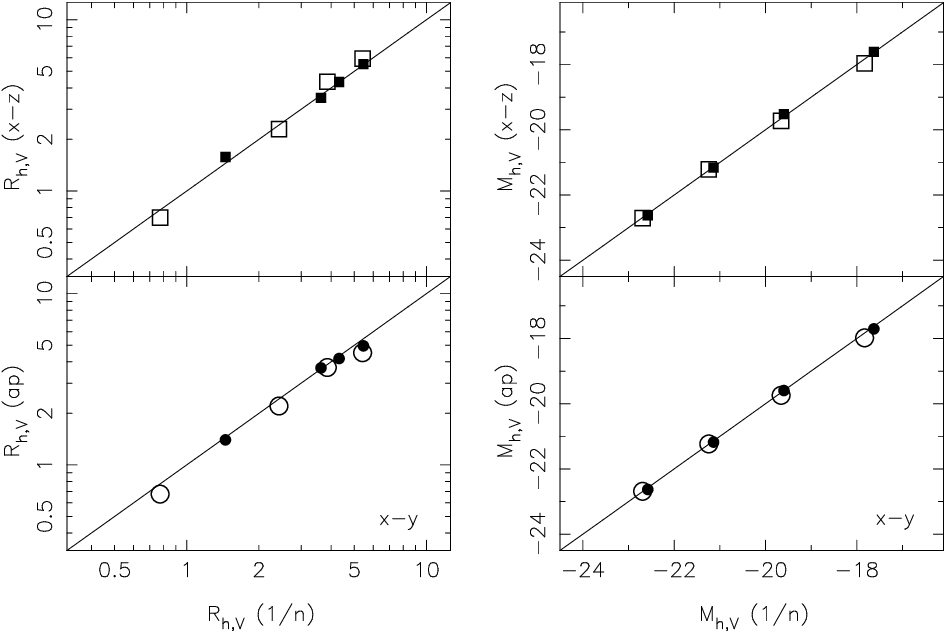}}
\figcaption[f3.eps]
{
 Comparison of the effective radius (left) and total magnitude (right)
 derived by different methods.
 The upper panels compare the value derived from the
 $r^{1/n}$ law fitting in the $x$-$y$ projection with that in the
 $x$-$z$ projection.  The lower panels compare the value
 derived from the $r^{1/n}$ law fitting (square) with the value
 derived from the large aperture photometry (circles)
 in the $x$-$y$ projection (see text for details).
 The open (solid) symbols indicate the value for model A (B).
\label{fa2cmre}}
\end{center}}

 We show the surface-brightness profiles and the color, metallicity,
and age gradients for model B2 in the right panels of Figures \ref{fa2sfb} and
\ref{fa2grad} in order to evaluate the effect of the difference in
the strength of the SNe feedback on these profiles and gradients. 
We analyzed these profiles and gradients also for the other models.
Table \ref{tabglp} summarizes the results.
In Table \ref{tabglp}, we can see that the strong feedback of model B
leads to a small $n$ in surface-brightness profiles and
a shallow slope in color and metallicity gradients, compared to
the ones of model A.
Moreover, with decreasing mass of the system, $n$ decreases
and the color and metallicity gradients becomes shallower in model B.
In other words, the strong feedback affects these profiles and gradients
more severely in lower mass systems.
Similar results are obtained in \citet{rc84a,myn99}.
The mass dependence of $n$ is consistent with the 
correlation between $n$ and luminosity in the observed
elliptical galaxies qualitatively \citep{ccd93}. On the other hand,
there is no correlation between the observed color and metallicity gradients
and the luminosity in luminous elliptical galaxies
\citep[][and reference therein]{pdidc90,ka00}, and
in less luminous elliptical galaxies ($M_B>-19$) there is 
still a large uncertainty \citep[e.g.,][]{vvls88}.

 The global photometric properties, such as the total luminosity,
the colors, and the effective radius, are also obtained from the
projected image data.
In this paper, we focus on the CMR, the Kormendy relation, and
the [Mg/Fe]--magnitude relation.
Thus we have to analyze the total magnitude, the color, the
effective radius, and [Mg/Fe] using the projected images.
The total magnitude and the effective radius
can be derived in two different ways. One method is to
fit the surface brightness profile to the $r^{1/n}$ law
(eq.\ [\ref{eqserlaw}]).
This fitting formula directly provides the effective radius, and
the integration from the center to the effective radius
gives half the total luminosity.
The other method is to assume that the aperture photometry
with a sufficiently large aperture provides the total value.
We assume that the luminosity derived with the 99 kpc aperture
is the total luminosity. Then the effective radius is defined
as the radius within which half the total luminosity is contained,
and calculated by counting the luminosity from the center
in the projected images. Throughout this paper
we set the center to the position of a pixel
which has the maximum $V$ band luminosity.
Figure \ref{fa2cmre} compares the effective radius and the total
magnitude derived by the different methods.
These panels show that there is no systematic difference.
Moreover Figure \ref{fa2cmre} indicates that the
total magnitude and the effective radius do not depend on the
direction of the projection.
It means that the end-products are nearly spherical
and the luminosity profiles are well approximated by the
$r^{1/n}$ law (see also Figs.\ \ref{fa2sfb}).
Finally, we use the former method in the following discussion.
Table \ref{tabglp} shows the total magnitude in the $B$, $V$, and $K$ bands and
the effective radius in the $V$ and $K$ bands for each model.
These data are derived from the fitting of the luminosity profiles
in the $x$-$y$ projections and used in the next section.

 In observation of elliptical galaxies,
colors and [Mg/Fe] are usually obtained using a fixed aperture.
Table \ref{tabappp} shows
the $U-V$ and $V-K$ colors and [Mg/Fe] for each model.
In the next section, we will compare the colors and [Mg/Fe]
of the simulation end-product in each model
with the observed values measured by BLE92a
and \citet{tfwg00a}. The values in Table \ref{tabappp}
have been obtained by the analysis using the same aperture as BLE92a
and \citet{tfwg00a}.
We made the projected image also for the luminosity weighted
abundance of Mg and Fe, and analyzed the ratio of the mean abundance 
of Mg to that of Fe within the aperture.

\vbox{
\begin{center}
\leavevmode
\hbox{
\epsfxsize=7cm
\epsffile{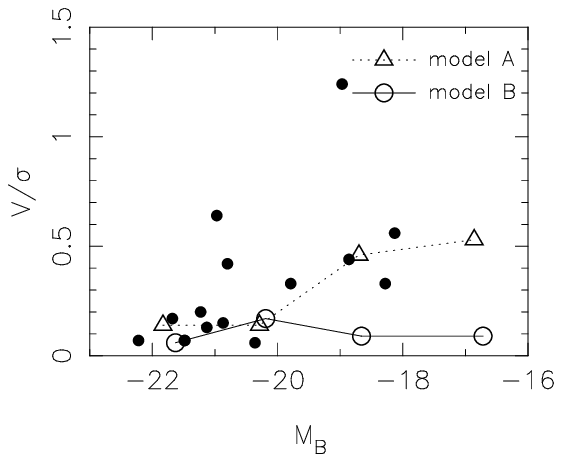}}
\figcaption[f4.eps]
{
 The comparison of $V/\sigma$ for the simulation end-products
and for galaxies in the Virgo and Coma clusters \citep[small dots,][]{def83}.
The triangles (circles) connected by
dotted (solid) lines indicate the $V/\sigma$ of model A (B).
\label{fvsig}}
\end{center}}

Due to calculating dynamical evolution, we can analyze
the dynamical properties
of end-products. K01 showed that the elliptical galaxy formation
model used in this paper leads to a small ratio of the rotational
velocity to the velocity dispersion ($V/\sigma$), which is consistent
with the observed $V/\sigma$ in bright elliptical galaxies.
The observed $V/\sigma$ in elliptical galaxies 
increases with decreasing luminosity \citep{def83}.
Now we consider the galaxies with different masses,
we analyzed $V/\sigma$ for all the models,
to examine the mass dependence of $V/\sigma$.
Figure \ref{fvsig} shows $V/\sigma$ against the B band absolute
magnitude. In model A, $V/\sigma$ increases slightly
with decreasing luminosity, like the tendency of the observational data.
On the other hand, the strong feedback of model B seems to 
lead to a low $V/\sigma$, compared to the one of model A, in low mass systems.
However, $V/\sigma$ of simulation end-products could not
be measured accurately, because their rotation curves do not provide
a clear flat shape, owing to their slow rotation and a poor spatial
resolution of numerical simulation. In other words, $V/\sigma$ indicated in
Figure \ref{fvsig} has ambiguity.
Thus we conclude that all the models provides a small $V/\sigma$
around $0.1\sim0.5$ irrespective of the luminosity or the feedback model. 
Our model cannot explain the existence of
less luminous elliptical galaxies with a large $V/\sigma$, especially
in model B, tough the observed $(V/\sigma)$s for less luminous galaxies have
a broad range.
We suppose that these systems might be formed from another initial condition,
e.g., with a larger $\lambda$ or a smaller $\sigma_{\rm 8, in}$ (K01).
This analysis made it clear that the final stellar systems
which we focus on have a small $V/\sigma$ 
and are supported mainly by random motions in terms of kinetics.
In the following sections, we concentrate on
the photometric properties.

\section{Global Scaling Relation}
\label{sgsrs}

 The photometric properties of the simulation end-products at $z=0$
which are derived by the data analysis described in Section \ref{sdataa}
are compared with the observed global scaling relations of
elliptical galaxies in this section.
We first compare the simulation
results with the observational data in the color versus magnitude
diagram. Next we discuss the sizes of the simulation end-products,
comparing them with the observed Kormendy relation.
Finally we examine the correlation between the total magnitude and
the abundance ratio of Mg to Fe, i.e., the [Mg/Fe]--magnitude relation,
which is sensitive to the star formation history.

\vbox{
\begin{center}
\leavevmode
\hbox{
\epsfxsize=7cm
\epsffile{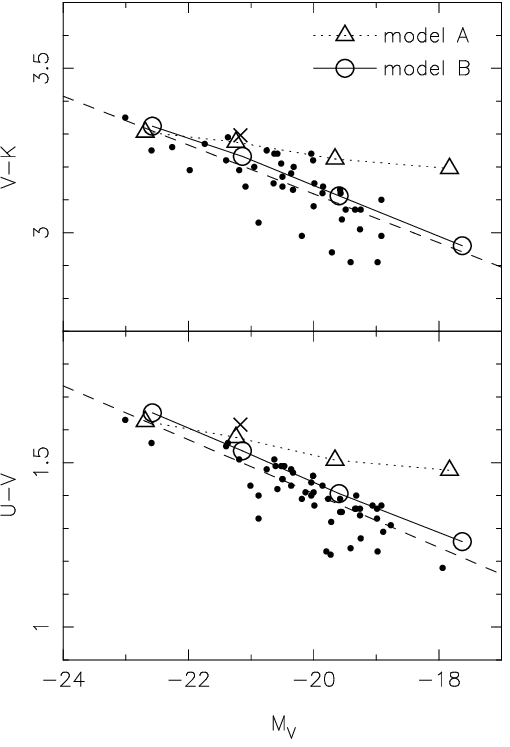}}
\figcaption[f5.eps]
{
 The comparison of the CMRs for the simulation end-products
 and the Coma cluster galaxies (small dots, BLE92a)
 in the aperture of 5 kpc. The triangles (circles) connected by
 dotted (solid) lines indicate the CMRs of model A (B). The dashed line
 shows the CMR fitted to the Coma cluster galaxies (BLE92b).
 The cross denotes the results of the same model as model A2 but with $c_*=1$
 (see Section \ref{ssf}).
\label{fcmr} }
\end{center}}

\subsection{The Color--Magnitude Relation}
\label{scmr}

 We construct the CMRs in model A and B.
Comparing model A with model B, we study the effect of the difference in
the strength of the SNe feedback.
Figure \ref{fcmr} shows the comparison of 
the simulation end-products and the observed galaxies in the Coma cluster
in the $V-K$ and $U-V$ CMR.
The data for galaxies in the Coma cluster are the observed CMR of BLE92a.
Since there is no difference between S0s and ellipticals
in the scaling relations which we discuss,
we do not distinguish S0s from ellipticals.
BLE92a supplies the $U-V$ and $V-K$ colors
which refer to an aperture size of 11 arcsec and
the $V$ band total magnitude derived from a combination of
their data and the literature.
Throughout this paper we adopt the distance modulus of the Coma
cluster of $m-M=34.7$ mag;
the Virgo distance modulus is $m-M=31.01$ \citep{gffkm99}
and the relative distance modulus of the Coma with respect to the Virgo
is $m-M=3.69$ (BLE92b).
This gives the luminosity distance of 87.1 Mpc for the Coma.
We assume that the angular diameter distance equals
the luminosity distance, because the redshift of the Coma
cluster ($z\sim0.023$) is nearly zero cosmologically.
Then the aperture size of 11 arcsec at the distance of the Coma cluster
corresponds to $\sim$ 5 kpc. 
The data for simulation end-products are tabulated in Tables
\ref{tabglp} and \ref{tabappp}.
Although model A has a tendency that smaller galaxies
have a bluer color, the slope is too shallow to
reproduce the observed slope of the CMR in the Coma cluster.
On the other hand, the slope of model B is steeper than 
that of model A, and model B well reproduces
not only the slope but also the zero-point of the observed CMR.
The success of model B is owing to the action of mass-dependent
galactic wind as shown later in detail.

\vbox{
\begin{center}
\leavevmode
\hbox{
\epsfxsize=7cm
\epsffile{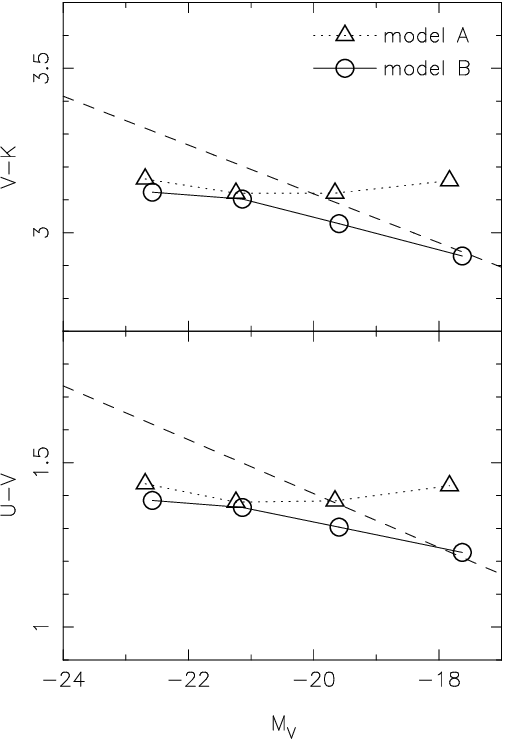}}
\figcaption[f6.eps]
{ The CMRs for the simulation end-products
 in the 99 kpc aperture. The triangles (circles) connected
 by dotted (solid) lines indicate the results of model A (B).
 The dashed lines are the same as that in Figure \ref{fcmr}.
\label{flacmr} }
\end{center}}

The slope of model A is just an aperture effect.
Elliptical galaxies have a color gradient
that the color at the center is redder than that in
the outer regions \citep[e.g.,][]{pdidc90}
and the end-products in our simulations also have gradients
(Fig.\ \ref{fa2grad} in Section \ref{sdataa}).
Thus the color within a fixed aperture provides
the color in a more central region for larger galaxies,
and the color of large galaxies becomes redder than that of small galaxies
even if the mean color of the whole galaxy is the 
same between the large and small galaxies.
Figure \ref{flacmr} shows the CMR for the mean color
within the aperture of 99 kpc, which covers almost the whole galaxy
in all the models which we examined.
The mean color of the whole galaxy in model A stays nearly constant
when the mass varies.
Hence the slope of model A in Figure \ref{fcmr} is caused only by the
smallness of the aperture size.
On the other hand, model B shows a significant slope
even when a large aperture size is specified as shown in Figure \ref{flacmr}.
Aperture effect cannot be ignored when we discuss the CMR
observed in an aperture of a small size.
Unfortunately, accurate colors evaluated within
various apertures have never been provided from any observations.
Recently \citet{ms01} showed that
the CMR derived using color measurements within the effective radius
is significantly flatter than those based on fixed-aperture color
measurements. Unfortunately, the photometric accuracy of 
their data was not good enough to derive an accurate CMR.
We would like to stress that an observation in apertures of various sizes
is quite important in understanding not only the CMR but also
the color gradients in elliptical galaxies.

\vbox{
\begin{center}
\leavevmode
\hbox{
\epsfxsize=7cm
\epsffile{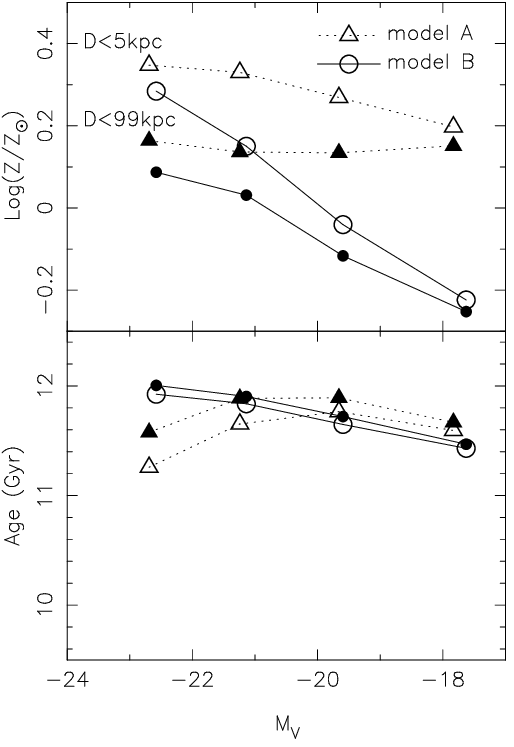}}
\figcaption[f7.eps]
{
  The metallicities (upper panel) and ages (lower panel)
 against the absolute $V$ band magnitude for each model.
 The triangles (circles) connected by dotted (solid) lines
 indicate the results of model A (B).
 The open (solid) symbols denote the values evaluated in the 5 kpc
 (99 kpc, spread over almost the whole galaxy) aperture.
\label{fzage} }
\end{center}}

 There are two equally plausible interpretations of the CMR \citep{gw94}.
The sequences of colors and line strengths among
elliptical galaxies can almost equally well be attributed to either age
difference or metallicity difference.
In the simulation this degeneracy can be broken completely,
because we can directly analyze the metallicity and age in the simulation
end-products. Figure \ref{fzage} shows the metallicity and age
for each model. Here the metallicity and age mean the luminosity weighted
value.  The age is very old irrespective of the models and the aperture.
The metallicity slope exhibits a similar behavior to the color
as a function of the luminosity.
Model A has a weak slope in the metallicity--magnitude diagram
of Figure \ref{fzage} when the aperture size is 5 kpc,
though the slope almost vanishes when the 99 kpc aperture is used.
On the other hand, model B shows a significant slope irrespective of
the aperture size, although a large aperture size leads to a shallower slope.
Thus it is concluded that the slope of CMR in Figure \ref{fcmr}
is caused solely by the effect of the metallicity.

\vbox{
\begin{center}
\leavevmode
\hbox{
\epsfxsize=8.5cm
\epsffile{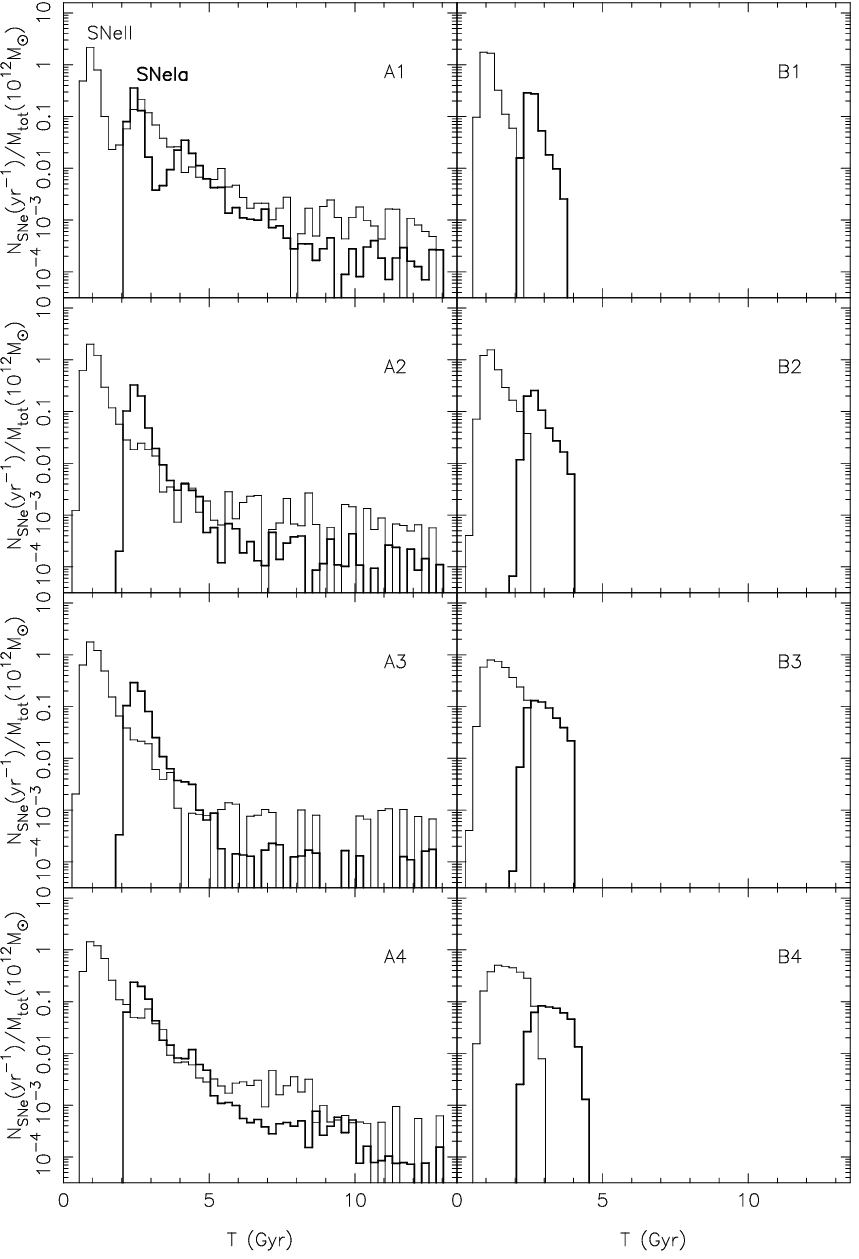}}
\figcaption[f8.eps]
{
 Time variations of the event rate of
 SNe II (normal lines) and SNe Ia (thick lines) for all the models.
\label{fsnh} }
\end{center}}

\begin{figure*}[t]
\epsfxsize=18.5cm
\epsffile{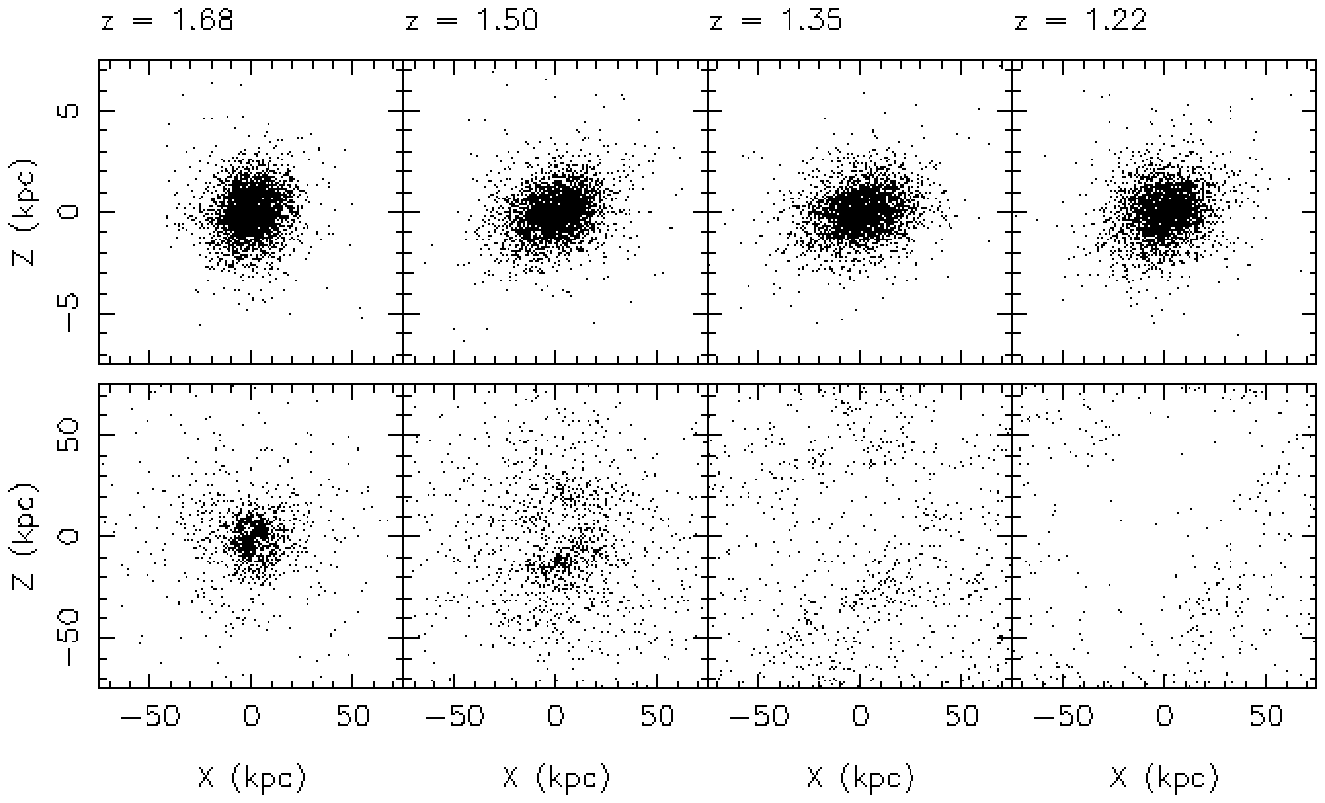}
\figcaption[f9s.eps]
{
 Time evolution of the system in model B4. The upper and lower panels
 shows the distributions of stellar particles and gas particles
 respectively. 
 Upper (lower) panels show the $x$-$z$ projection of the particles and
 measure 15 (150) kpc across.
\label{fb4anim} }
\end{figure*}

 As a result, model B succeeds in reproducing the CMR by causing a 
decrease in the metallicity of the stellar contents
for low-mass systems. On the other hand, model A completely fails.
To see the origin of this difference, we examine
the star formation history for each model. Figure \ref{fsnh}
shows the histories of SNe II and SNe Ia event rates normalized 
by the total mass of the system.
In our simulations, the histories of SNe II events completely trace
those of star formation, because we assume the instantaneous
recycling (see Section \ref{ssf} for details).
On the other hand, SNe Ia occur 1.5 Gyr behind SNe II.
We can clearly see that
in model B star formation abruptly ceases around $t=2.5\sim3$ Gyr 
($z=2\sim1.7$).
The dynamical evolution of the system clearly demonstrates that
this is caused by the so-called galactic wind. Figure \ref{fb4anim}
which shows the morphological evolution of model B4.
Around $z\sim1.7$, gas particles begin to blow out from
the stellar system, i.e., the galactic wind occurs.
All the gas particles overcome
the binding energy of the dark matter and stars, and escape from the system.
In addition we found that the mass fraction of the ejected gas increases with
decreasing mass of the system in model B, as seen in Figure \ref{fgwm}.
The ejected gas mass in Figure \ref{fgwm} is defined as the total mass of
all the gas particles whose galactocentoric radius is greater than 20 kpc.
Since the ejected gas cannot contribute to the further
metal enrichment, a higher mass system suffers from more enhancement
in heavy elements. This mass dependence of the mass fraction of the 
gas ejected by the galactic wind
is the origin of the metallicity--magnitude relation shown 
in Figure \ref{fzage}, which in turn causes the CMR shown in Figure \ref{fcmr}.
There is little difference models B1 and B2 in Figure \ref{fgwm}.
In this luminosity range, 
the aperture effect is dominant factor in explaining the
observe slope. However, the metallicity--magnitude relation
caused by the galactic wind is essential in order to reproduce
the global CMR in the observed luminosity range.

 It is notable that the galactic wind occurs immediately
after SNe Ia ignite. It is clearly shown that SNe Ia are important
for the galactic wind. The event rate of SNe Ia exceeds that of SNe II
after the peak of star formation. SNe Ia occur when
the gas is exhausted by star formation
and the gas density becomes low.
Thus SNe Ia can strongly affect the evolution of the residual gas. 

On the other hand,
Figure \ref{fgwm} shows that almost all the gas is exhausted by $z=0$
irrespective of the mass of the system in model A.
As a consequence, model A fails to reproduce the observed CMR.
The difference in results of models A and B
is the strength of the effect of SNe.
As a result, a strong effect of the SNe feedback like the one included
in model B is required to explain the observed CMR.
Moreover we can clearly see from the results of model B that
the mass fraction of the gas ejected by the galactic wind
depends on the mass of the system and that SNe Ia play
a quite important role in the outbreak of galactic wind.

\vbox{
\begin{center}
\leavevmode
\epsfxsize=7cm
\hbox{
\epsfxsize=7cm
\epsffile{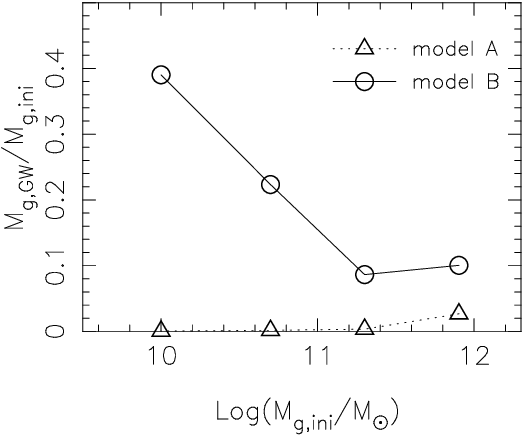}}
\figcaption[f10.eps]
{ The ratio of the ejected gas mass ($M_{\rm g,GW}$)
  at $z=0$ to the initial gas mass ($M_{\rm g,ini}$).
 The triangles (circles) connected by dotted (solid) lines
 indicate the data of model A (B). Here, the ejected gas
 is defined as the sum of all the gas particles whose
 galactocentoric radius is greater than 20 kpc.
\label{fgwm} }
\end{center}}

\subsection{The Kormendy Relation}
\label{skr}

 Numerical simulations have a strong advantage
over pure chemical evolution studies and semi-analytic models
of galaxy formation
in that simulations can provide information about the structure
of the end-products. Owing to this advantage,
we can discuss the scaling relation
which prescribes the size of the elliptical galaxies,
i.e., the Kormendy relation.
Figure \ref{fmre} displays the comparison of the Kormendy relations
for the simulation end-products and the Coma cluster galaxies
both in the $V$ and $K$ bands.
We refer to the data of the Coma cluster galaxies of \citet{mp99}.
When we derive the absolute magnitude and the effective
radius in the kpc unit from the data set in \citet{mp99},
we assume the same distance modulus as mentioned above.
In Figure \ref{fmre}, both models have a tendency that higher mass galaxies
have larger effective radii, which is qualitatively consistent with
the Kormendy relation for the observed elliptical galaxies.
However the slope is too shallow in both models to
consistent with the observed relation quantitatively.
In particular, the slope of model B is much shallower than that of
model A. This is due to the expansion of the low mass systems
induced by the strong feedback.
It is concluded that
the strong feedback which is required to explain the 
observed CMR is unsuitable for explaining the Kormendy relation.
The failure in reproducing the Kormendy relation
poses a serious problem to the galactic wind scenario.

\vbox{
\begin{center}
\leavevmode
\epsfxsize=7cm
\hbox{
\epsfxsize=7cm
\epsffile{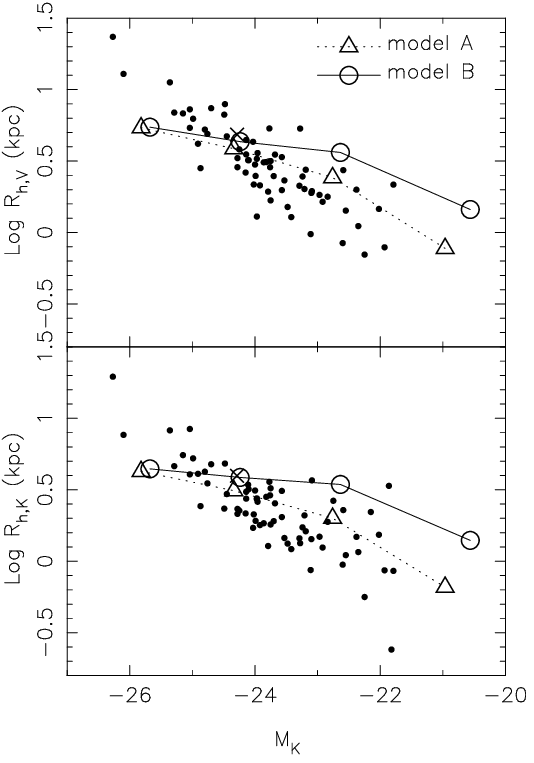}}
\figcaption[f11.eps]
{
 The comparison of the Kormendy relations for the simulation end-products
 and the Coma cluster galaxies \citep[small dots,][]{mp99} in the $V$
 (upper panel) and $K$ (lower panel) bands.
  The triangles (circles) connected by
 dotted (solid) lines indicate the data for model A (B). 
 The cross denotes the results of the same model as model A2 but with $c_*=1$
 (see Section \ref{ssf}).
\label{fmre} }
\end{center}}

\subsection{The [Mg/Fe]--Magnitude relation}
\label{smgfe}

 Our numerical code takes account of both SNe II and SNe Ia,
and follows the evolution of the abundances of
several chemical elements (C, O, Ne, Mg, Si, and Fe).
Therefore we can examine the abundance ratio of chemical elements
in the simulation end-products.
Observational data indicate that [Mg/Fe] correlates with
the luminosity, i.e., the [Mg/Fe]--magnitude relation (see also Section
\ref{sintro}). We compare the [Mg/Fe]--magnitude relation for the
simulation end-products with the observational data
of local field and group elliptical galaxies.
We use the observational data of [Mg/Fe]
in \citet{tfwg00a} and the absolute B band magnitudes in
\citet{tfwg00b}. 
\citet{tfwg00a} derived [Mg/Fe] in the $r_{e,B}/2$ and $r_{e,B}/8$ apertures,
using an extension of the \citet{gw94} models that incorporates nonsolar
line-strength response functions by \citet{tb95}.
We use the data for model 4 of \citet{tfwg00a},
which is the most successful model according to their paper.
In Figure \ref{fmgfe} we compared the $r_{e,B}/2$ aperture data 
with [Mg/Fe] for the simulation end-products in the  $r_{e,B}/2$ aperture. 
Since all the models do not reproduce 
the observed effective radius as shown in Section \ref{skr},
the physical aperture size in all the models is not consistent
with that applied in the observed galaxies with the same luminosity.
Fortunately, the observed gradient of [Mg/Fe]
is quite weak in these samples according to \citet{tfwg00a}.
On the other hand, Figure \ref{fa2grad} shows the [Mg/Fe] gradients
for models A2 and B2. Although there is a little gradient of [Mg/Fe]
in the simulation results of model A, the gradient becomes shallower within 
the effective radius irrespective of models.
Thus the difference in the aperture size is not so important in this case.
The observed data show a clear tendency
that [Mg/Fe] increases with the galactic luminosity.
However both models A and B are incapable of reproducing this tendency.
In Figure \ref{fsnh} the star formation histories 
in both models A and B
are nearly identical for the whole mass range of the system,
although there is a clear difference between models A and B.
Consequently, [Mg/Fe] is constant irrespective of the luminosity of
the system.  In addition [Mg/Fe] in model B is systematically
higher than that in model A. The reason is that in all of model B
the galactic wind leads to the cessation of star formation
before the metal enrichment by SNe Ia progresses,
while in all of model A 
star formation continues after SNe Ia ignite.

 Model B provides a little different result in [Mg/Fe]--magnitude
diagram from what pure chemical evolution studies 
of the galactic wind scenario predicted \citep[e.g.,][]{fm94}. 
In pure chemical evolution studies, 
the efficiency of star formation is assumed to be constant
irrespective of the mass of the systems or
an increasing function with decreasing mass.
Therefore, the galactic wind in higher mass systems need to occur
later than that in lower mass systems in order to explain the observed CMR,
and the duration of star formation is longer
in the higher mass systems than that in the lower ones.
The elements of Mg and Fe are mostly produced by SNe II and
SNe Ia respectively, and SNe Ia have a longer delay
than SNe II with respect to the formation of stars.
A galaxy with a longer time duration of star formation
is much more enriched by SNe Ia and gets a lower [Mg/Fe].
Hence pure chemical evolution studies predict that
[Mg/Fe] is a decreasing function of the
galactic mass and luminosity, which provides an opposite
slope to the observation \citep{mt87}.
On the other hand, in model B [Mg/Fe] is constant
irrespective of the luminosity.
Comparison of the peak SNe II event rate for each model
in Figure \ref{fsnh} shows that the efficiency of star formation
increases with mass of the system. It means that the strong feedback
causes a self-regulation of star formation in lower mass systems
\citep[see also][]{lcs00}.
Consequently, the galactic wind in all the models occurs
at nearly the same time, though the higher mass systems converts
a larger fraction of gas into stars as seen in Figure \ref{fgwm}.
Therefore, our numerical simulation which calculates the dynamical
evolution self-consistently reveals that 
the mass-dependent self-regulation of star formation
is caused by the strong feedback, so that
the problem of the galactic wind scenario
in the [Mg/Fe]--magnitude relation is improved. 
Finally, the high constant [Mg/Fe] in model B is caused by
both effects of the cessation of star formation in an early epoch
irrespective of the mass of systems
and the mass-dependent self-regulation of star formation, which are
induced by the feedback strong enough to reproduce the observed CMR.
However, the observed slope is still not reproduced completely.

\vbox{
\begin{center}
\leavevmode
\epsfxsize=7cm
\hbox{
\epsfxsize=7cm
\epsffile{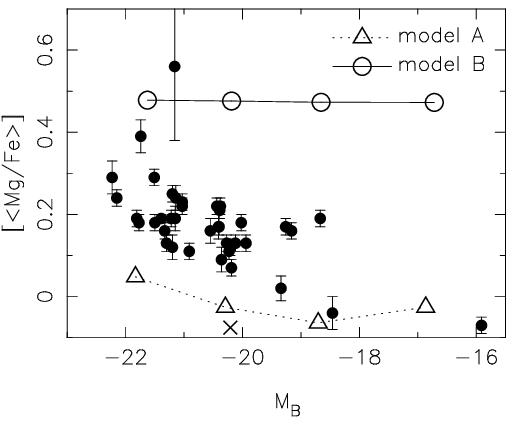}}
\figcaption[f12.eps]
{
 The comparison of the simulation end-products with the early-type
 galaxies in the observed [Mg/Fe] \citep{tfwg00a} vs.\ $M_B$ diagram.
 The triangles (open circles) connected by dotted (solid) lines
 indicate the data for model A (B).
 The cross denotes the results of the same model as model A2 but with $c_*=1$
 (see Section \ref{ssf}).
\label{fmgfe} }
\end{center}}

\section{Discussion and Conclusions}
\label{sdisc}

 We have numerically studied the dynamical and chemical processes
of the formation of elliptical galaxies in the CDM universe,
in order to examine the origin of the mass-dependence of
the photometric properties of elliptical galaxies.
Our numerical simulations take account of both SNe II and SNe Ia,
and follow the time evolution of the abundances of several chemical
elements (C, O, Ne, Mg, Si, and Fe).
In this paper we paid a special attention to the CMR of elliptical
galaxies. In addition we examined the Kormendy relation,
which prescribes the size of elliptical galaxies, and
the [Mg/Fe]--magnitude relation, which provides a strong constraint
on the star formation history.
Because the SNe feedback is likely to be a crucial determinant of the
properties of the end-products, we compared two different
strengths of the SNe feedback, the minimum (model A) and strong
(model B) feedback. 
Although our feedback implementations are very crude,
we could examine whether or not the SNe feedback should strongly affect
the evolution of the system to explain the global
scaling relations of elliptical galaxies.

 In conclusion, the strong influence of SNe like the one adopted
in model B is required to reproduce the observed CMR
as shown in Figure \ref{fcmr}.
We found that the feedback affects the evolution of lower mass
systems more strongly, so that a larger fraction of gas is blown out in
a lower mass system (Figure \ref{fgwm}).  
Consequently, higher mass systems become more metal rich and
have redder colors than lower mass systems (Figure \ref{fzage}).
These results are consistent with the predictions of 
pure chemical evolution studies based on the galactic wind scenario
\citep[e.g.,][]{ay87}.
The mass dependence of the metal enrichment has already been examined
qualitatively also in the realistic numerical studies.
However, previous authors adopted a simple model
of elliptical galaxy formation in which an elliptical galaxy
is formed by the monolithic collapse of a single homogeneous
proto-galactic cloud
\citep[e.g.,][]{rc84a,rc84b,ft98,myn99}
or the mergers of two disk galaxies \citep[e.g.,][]{bs98, bs99}.
We adopted the scenario that
the elliptical galaxies are formed through the clustering
of the sub-clumps caused by the initial small-scale density perturbations.
This modeling is based on the CDM cosmology and
has a reasonable cosmological ground.
Thus the present model is more realistic than those used by previous authors.
We showed that the mass-dependence of metal enrichment
is naturally induced in this scenario. The present work demonstrates that
many conclusions of pure chemical evolution studies 
based on the monolithic collapse picture are valid also in
elliptical galaxy formation through the hierarchical clustering.

 The assumed energy of each SN ($4\times10^{51}$ ergs) in model B 
might seem to be too strong.
It might suggest that another mechanism ignored in this study is needed
in order to enhance the effect of the SNe feedback on the evolution of systems.
For example, the UV background radiation is a plausible candidate,
because it is possible to suppress the cooling in 
low mass systems, as discussed later.

 We were able to evaluate relative importance of two types of SNe,
SNe II and SNe Ia. SNe II occur immediately after star formation events,
while there is a delay of 1.5 Gyr between star formation and SNe Ia.
Therefore, SNe II occur in dense gaseous environments still actively forming
stars, whereas SNe Ia act on tenuous gas left after star formation.
This difference makes SNe Ia the major trigger of the galactic wind.
This result is a novel one obtained from our numerical simulations which
calculate gas dynamics, radiative cooling, star formation, and
the SNe II and SNe Ia feedback self-consistently, and
means that the energy feedback of SNe Ia is of fundamental importance
for the evolution of elliptical galaxies.
However, it should be noted that our SNe modeling is very simple.
As mentioned in Section \ref{ssneia}, we neglect the spread of the
lifetime of SNe Ia and assume that all SNe Ia explode 1.5 Gyr after
star formation. This assumption might cause too great
impact on the interstellar medium.
Modeling of SNe II is also simplified by an assumption of 
the instantaneous recycling. Our average time step was $\sim 1.7\times10^5$
yr. It is considerably shorter than the typical lifetime of SNe II progenitors.
The effects of feedback of SNe II might also be overestimated.
 In order to compare the effects of SNe II and SNe Ia
quantitatively, it is required to adopt more sophisticated models of SNe Ia
and SNe II which consider the realistic spread of lifetimes of SNe Ia
and SNe II progenitors. We will address this issue in a forthcoming paper.
Nevertheless, our results are still important in that
it is clearly shown that the delayed SNe feedback, such as SNe Ia,
affects the dynamical evolution of interstellar medium more efficiently 
than the instantaneous one.

 In Table \ref{tabsnr} we compare the SNe II and Ia rates of
the simulation end-products at $z=0$ to the observed rates 
at nearby galaxies \citep{cet99}. The present SNe rates 
are obtained by the average of the number of SNe which occurred
within 1.5 Gyr and 2 Gyr. These values slightly depend on 
the duration of the sampling in model A, as expected from Figure \ref{fsnh}.
In all the models of model A SNe II rate is too high to reproduce
the observed rate, which suggests few or no SNe II.
Although model B is in agreement with no SNe II,
the SNe Ia rate is inconsistent with the observation, which
suggests a non-zero event rate.
We anticipate that SNe Ia model which takes into account the spread lifetime
of SNe Ia progenitors would eliminate disagreement between these SNe Ia rates,
because the continuous SNe Ia is expected after star formation ceased.

 In analyzing the colors of simulation end-products,
we payed attention to the aperture effect
by taking the same aperture size as adopted in the actual observations.
The color evaluated in an aperture
fixed to a small size leads to a non-vanishing slope in
the color--magnitude diagram, even if the sample galaxies have
an identical mean color (i.e., the color averaged over the
entire galaxy) as seen in model A.
This is because elliptical galaxies generally have
color gradients such that the center of the galaxy is redder than
the outer regions.
Hence we would like to stress that the aperture effect should not be ignored
when we discuss the CMR observed in an aperture fixed to a small size.

 Our numerical models failed to quantitatively reproduce
the Kormendy relation, which prescribes the size of elliptical galaxies. 
Both models A and B give a significantly shallower slope than the observed one.
Especially, model B, which can reproduce the observed CMR,
causes too large effective radii in low mass systems due to
the expansion of the system induced by the strong feedback.
It means that the feedback strong enough to explain the observed CMR
leads to not only the galactic wind but also the expansion of the system.
This result rises a serious problem to the galactic wind scenario.

 The diagram of the abundance ratio [Mg/Fe] versus
the galactic luminosity provides an important diagnostic to infer
the star formation history in galaxies, due to the different
timescales for the Mg and Fe production.
The observed [Mg/Fe]--magnitude relation indicates that
more luminous galaxies have larger [Mg/Fe].
However, both models A and B leads to nearly constant [Mg/Fe]
as a function of the mass, and [Mg/Fe] of model B is substantially
larger than that of model A due to the short duration of star formation.
Both of models thus failed to reproduce
the observed [Mg/Fe]--magnitude relation.
However, we must bear in mind that 
theoretical prediction of [Mg/Fe] still remains ambiguous,
because the ratio depends on some unknown physical processes,
such as the lifetime of the progenitor of SNe Ia
and the nucleosynthesis yields of SNe II and SNe Ia.

This discrepancy in the [Mg/Fe]--magnitude relation
has already been known as a problem of the galactic
wind scenario. As mentioned in Section \ref{smgfe},
our numerical simulation in the strong feedback model, i.e., model B, 
reduces the discrepancy, compared to the prediction of
pure chemical evolution studies, due to the mass-dependent
self-regulation of star formation.
However, to explain the observed trend quantitatively, 
we still have to consider 
what physical process ignored in the present study can
lead to a small [Mg/Fe] in lower mass systems.
\citet{fwg92} proposed
three alternative scenarios to solve this discrepancy:
i) different star formation timescales; ii)
a selective loss of Mg and Fe; and iii) a selective production of
Mg vs.\ Fe, i.e., variable IMF.
There is no physical reason yet why the IMF depends on
the mass of the system. It is a natural consideration that
the IMF depends on the local physical condition of the interstellar gas
rather than the mass of the host system. Thus the scenario iii) sounds
infeasible. In our numerical simulations, ii) did not occur.
As a physical process to realize i) we can consider 
the UV background radiation which we ignored in this paper.
The UV background radiation can suppress the cooling and
star formation more efficiently in lower mass galaxies
\citep{ge92}, which is also confirmed by realistic numerical
simulations \citep[e.g.,][]{qke96,tw96,whk97}.
Thus it is possible to decrease the efficiency of star formation
and prolong star formation with decreasing mass of the system.
A forthcoming paper will discuss the effect of UV background radiation on
the [Mg/Fe] in lower mass galaxies.

 The galaxy formation model used here parameterizes the properties
of the seed galaxy by the total mass, the spin parameter,
the amplitude of the small scale density perturbations,
and the collapse redshift, i.e., $M_{\rm tot}$, $\lambda$,
$\sigma_{8,\rm in}$, and $z_c$. This approach turned out to be quite useful
in studying the evolution of a single seed galaxy in detail.
In this paper we studied how the properties of the end-product
depend on the total mass of the system, $M_{\rm tot}$,
when other parameters are fixed.
However, little is known about whether and how the above parameters depend on
each other.
On the other hand, in cosmological simulations,
the above parameters which are assumed to be
independent of each other in this paper are automatically 
determined in the process of structure formation in the universe.
Hence, a cosmological simulation
which is able to spatially resolve the process of the formation and
evolution of individual galaxies is indispensable in studying the
connection between the evolution of galaxies and that of the universe.
In future work we intend to investigate what is the origin of the observed
scaling relations of the elliptical galaxies,
i.e., what controls those relations,
using a high-resolution cosmological simulation.

\acknowledgments

D.K.\ would like to thank Masafumi Noguchi for invaluable discussion.
D.K.\ also thanks Brad K. Gibson for helpful suggestions.
D.K.\ is grateful to Nobuo Arimoto and Tadayuki Kodama
for kindly providing the tables of their SSPs data.
D.K.\ is also grateful to Edmund Bertschinger for generously providing
the COSMICS programs.
D.K.\ acknowledges the Yukawa Institute Computer Facility
and the Astronomical Data Analysis Center of the National Astronomical
Observatory, Japan
where the numerical computations for this paper were performed.
This work was supported in part by
the Japan Atomic Energy Research Institute.



\begin{deluxetable}{llll}
\tablecolumns{4}
\tablewidth{0pc} 
\tablecaption{Nucleosynthesis products of SNe II and SNe Ia
 \label{tblmet-1} }
\tablehead{
\colhead{}
 & \multicolumn{2}{c}{Synthesized mass\tablenotemark{a} ($\rm M_\odot$)} &
  \colhead{} \\
\colhead{Element} & \colhead{Type II ($M_{i, {\rm SNe II}}$)} &
\colhead{Type Ia ($M_{i, {\rm SNe Ia}}$)} &
\colhead{Solar abundance\tablenotemark{b}}
} 
\startdata 
 $^{12}$C & $7.91\times10^{-4}$ & $7.32\times10^{-5}$ &
  $0.30\times10^{-2}$ \\
 $^{16}$O & $2.01\times10^{-2}$ & $2.17\times10^{-4}$
  & $0.96\times10^{-2}$ \\
 $^{20}$Ne & $2.24\times10^{-3}$ & $3.06\times10^{-6}$ &
  $0.16\times10^{-2}$ \\
 $^{24}$Mg & $9.38\times10^{-4}$ & $1.29\times10^{-5}$ &
  $0.52\times10^{-3}$ \\
 $^{28}$Si & $1.10\times10^{-3}$ & $2.27\times10^{-4}$ &
  $0.65\times10^{-3}$ \\
 $^{56}$Fe & $7.79\times10^{-4}$ & $9.29\times10^{-4}$ &
  $0.13\times10^{-2}$ \\
\enddata
\tablenotetext{a}{ The Salpeter IMF with $M_{\rm u}=60\ {\rm M_\odot}$ and
  $M_{\rm l}=0.4\ {\rm M_\odot}$}
\tablenotetext{b}{ Ferrini et al.\ 1992}
\end{deluxetable}


\begin{deluxetable}{ccccccccc}
\tablecolumns{9}
\tablewidth{0pc} 
\tablecaption{Model Parameters \label{tblmps}}
\tablehead{
\colhead{Model} & \colhead{$N_{\rm p}$} & \colhead{$M_{\rm tot}$} &
 \multicolumn{2}{c}{Particle Mass (${\rm M_\odot}$)} &
 \multicolumn{2}{c}{Softening (kpc)} & \colhead{$\epsilon_{SN}$} &
 \colhead{$f_v$} \\
 \colhead{Name} & \colhead{} & \colhead{($\rm M_\odot$)} &
 \colhead{DM} & \colhead{Gas} & \colhead{DM} & \colhead{Gas} &
 \colhead{} & \colhead{}
} 
\startdata 
A1 & $9171\times2$ &
 $8\times10^{12}$ & $7.85\times10^{8}$ & $8.72\times10^{7}$ &
 5.39 & 2.59 & 0.1 & 0 \\
A2 & $9171\times2$ &
 $2\times10^{12}$ & $1.96\times10^{8}$ & $2.18\times10^{7}$ &
 3.40 & 1.63 & 0.1 & 0 \\
A3 & $9171\times2$ &
 $5\times10^{11}$ & $4.91\times10^{7}$ & $5.45\times10^{6}$ &
 2.14 & 1.03 & 0.1 & 0 \\
A4 & $9171\times2$ &
 $1\times10^{11}$ & $9.81\times10^{6}$ & $1.09\times10^{6}$ &
 1.25 & 0.60 & 0.1 & 0 \\
B1 & $9171\times2$ &
 $8\times10^{12}$ & $7.85\times10^{8}$ & $8.72\times10^{7}$ &
 5.39 & 2.59 & 4 & 0.9 \\
B2 & $9171\times2$ &
 $2\times10^{12}$ & $1.96\times10^{8}$ & $2.18\times10^{7}$ &
 3.40 & 1.63 & 4 & 0.9 \\
B3 & $9171\times2$ &
 $5\times10^{11}$ & $4.91\times10^{7}$ & $5.45\times10^{6}$ &
 2.14 & 1.03 & 4 & 0.9 \\
B4 & $9171\times2$ &
 $1\times10^{11}$ & $9.81\times10^{6}$ & $1.09\times10^{6}$ &
 1.25 & 0.60 & 4 & 0.9 \\
\enddata
\end{deluxetable}


\begin{deluxetable}{ccccccccccccc} 
\tablecolumns{13}
\tablewidth{0pc} 
\tablecaption{Global photometric properties. \label{tabglp}}
\tablehead{
 \colhead{Model} & \colhead{} & \multicolumn{3}{c}{V band} & \colhead{}
 & \multicolumn{3}{c}{K band} & \colhead{} & \colhead{$\Delta(B-R)$} &
 \colhead{} & \colhead{$\Delta\log(Z/Z_\odot)$} \\
\cline{3-5} \cline{7-9} \cline{11-11} \cline{13-13}
 \colhead{Name} & \colhead{$M_B$} & \colhead{$M_V$} & \colhead{$n$} & \colhead{$r_e$} 
 & \colhead{} & \colhead{$M_K$} & \colhead{$n$} & \colhead{$r_e$} & \colhead{} &
 \colhead{$\Delta\log(r)$} & \colhead{} & \colhead{$\Delta \log(r)$} \\
 \colhead{(1)} & \colhead{(2)} & \colhead{(3)} & \colhead{(4)} &
 \colhead{(5)} & \colhead{} & \colhead{(6)} & \colhead{(7)} &
 \colhead{(8)}& \colhead{} & \colhead{(9)} & \colhead{} & \colhead{(10)} 
}
\startdata 
A1 & $-21.83$ & $-22.69$ & 4.18 & 5.41 & & $-25.82$ & 3.96 & 4.26 & &
 $-0.16$ & & $-0.38$ \\
A2 & $-20.29$ & $-21.24$ & 3.17 & 3.86 & & $-24.34$ & 3.06 & 3.12 & &
 $-0.20$ & & $-0.51$ \\
A3 & $-18.70$ & $-19.66$ & 3.41 & 2.42 & & $-22.76$ & 3.39 & 2.00 & &
 $-0.11$ & & $-0.40$\\
A4 & $-16.86$ & $-17.83$ & 3.51 & 0.77 & & $-20.96$ & 3.42 & 0.66 & &
 $-0.20$ & & $-0.52$ \\  
B1 & $-21.63$ & $-22.58$ & 2.12 & 5.45 & & $-25.82$ & 3.96 & 4.26 & &
 $-0.33$ & & $-0.63$ \\  
B2 & $-20.19$ & $-21.14$ & 1.41 & 4.32 & & $-24.24$ & 1.35 & 2.86 & &
 $-0.17$ & & $-0.26$ \\  
B3 & $-18.66$ & $-19.59$ & 0.92 & 3.63 & & $-22.64$ & 0.89 & 3.45 & &
 $-0.10$ & & $-0.15$ \\  
B4 & $-16.72$ & $-17.63$ & 0.96 & 1.45 & & $-20.56$ & 0.95 & 1.40 & &
 $-0.08$ & & $-0.11$ \\
\enddata
\tablecomments{Col.\ (10): Luminosity weighted metallicity.}
\end{deluxetable}


\begin{deluxetable}{ccccccccccc} 
\tablecolumns{11}
\tablewidth{0pc} 
\tablecaption{
 Photometric properties and stellar populations within apetures.
 \label{tabappp}}
\tablehead{
 \colhead{Model} &
 \multicolumn{4}{c}{$D<$5 kpc} & \colhead{} & 
 \multicolumn{4}{c}{$D<$99 kpc} & \colhead{}
  \\ \cline{2-5} \cline{7-10}
 \colhead{Name} & \colhead{$U-V$} & \colhead{$V-K$} & \colhead{[Z/H]}
 & \colhead{Age} & \colhead{} &
  \colhead{$U-V$} & \colhead{$V-K$} & \colhead{[Z/H]} & \colhead{Age} 
 & \colhead{[Mg/Fe]} \\
 \colhead{(1)} & \colhead{(2)} & \colhead{(3)} & \colhead{(4)} &
 \colhead{(5)} & \colhead{} & \colhead{(6)} & \colhead{(7)} &
 \colhead{(8)} & \colhead{(9)} & \colhead{(10)} 
}
\startdata 
A1 & 1.63 & 3.31 & 0.35 & 11.3 & & 1.35 & 3.10 & 0.16 & 11.6 & 0.05 \\
A2 & 1.58 & 3.27 & 0.33 & 11.7 & & 1.35 & 3.09 & 0.14 & 11.9 & $-0.03$ \\
A3 & 1.51 & 3.22 & 0.27 & 11.8 & & 1.34 & 3.09 & 0.13 & 11.9 & $-0.06$ \\
A4 & 1.48 & 3.19 & 0.20 & 11.6 & & 1.34 & 3.09 & 0.15 & 11.8 & $-0.03$ \\  
B1 & 1.65 & 3.32 & 0.28 & 11.9 & & 1.35 & 3.10 & 0.09 & 12.0 & 0.48 \\
B2 & 1.54 & 3.23 & 0.15 & 11.8 & & 1.32 & 3.08 & 0.03 & 11.9 & 0.48 \\
B3 & 1.41 & 3.11 & $-0.04$ & 11.6 & & 1.29 & 3.04 & $-0.12$ & 11.7 & 0.47 \\
B4 & 1.26 & 2.96 & $-0.22$ & 11.4 & & 1.22 & 3.01 & $-0.25$ & 11.5 & 0.47 \\
\enddata
\tablecomments{
 Col.\ (4) and (8): Luminosity weighted
 metallicity $\log (Z/Z_\odot)$.
 Col.\ (5) and (9): Luminosity weighted age (Gyr). 
 Col.\ (10): [Mg/Fe] in the $r_{e,B}/2$ aperture.
 }
\end{deluxetable}


\begin{deluxetable}{cccccc}
\tablecolumns{5}
\tablewidth{0pc} 
\tablecaption{ SN rates. Units are
 $SNu={\rm SN} (100 {\rm yr})^{-1} (10^{10} L_{\odot, B})$.
 \label{tabsnr}}
\tablehead{
 \colhead{} &
 \multicolumn{2}{c}{1.5 Gyr\tablenotemark{a}} & &
 \multicolumn{2}{c}{2 Gyr\tablenotemark{b}}
  \\ \cline{2-3} \cline{5-6}
 \colhead{Model} & \colhead{SNe Ia} & \colhead{SNe II} & 
 & \colhead{SNe Ia} & \colhead{SNe II} 
}
\startdata 
A1 & 0.18 & 0.47  & & 0.17 & 0.74 \\
A2 & 0.09 & 0.38  & & 0.13 & 0.37 \\
A3 & 0.10 & 0.43  & & 0.08 & 0.47 \\
A4 & 0.07 & 0.23  & & 0.08 & 0.31 \\
B1 & 0.00 & 0.00  & & 0.00 & 0.00 \\
B2 & 0.00 & 0.00  & & 0.00 & 0.00 \\
B3 & 0.00 & 0.00  & & 0.00 & 0.00 \\
B4 & 0.00 & 0.00  & & 0.00 & 0.00 \\
 & & & & & \\
E-S0\tablenotemark{c}
 & 0.18$\pm$0.06 & $<$0.02 & & & \\
\enddata
\tablenotetext{a(b)}
{ Avarage SNe rates for 1.5 (2) Gyr. }
\tablenotetext{c}{ Observed SN rate in \citet{cet99}.}
\end{deluxetable}

\end{document}